\documentstyle[12pt]{article}

\input defphys.inc

\def\vecn{\vec m}

\def\LSMO{La$_{1-x}$Sr$_x$MnO$_3$}

\catcode`\@=11
\newcount\@arga
\newcount\@argb
\newcount\@argc
\newcount\@ctmpa
\newcount\@ctmpb
\newcount\@ctmpc
\newcount\@ctmpd
\newcount\@ctmpe
\newdimen\@darg
\newdimen\@bblen
\newif\if@bbllx
\newif\if@bblly
\newif\if@bburx
\newif\if@bbury
\newif\if@height
\newif\if@width
\newif\if@scale
\newif\ifno@bb
\newif\ifepsfdraft
\epsfdraftfalse
\def\@setpsfile#1{
                \typeout{epsf:[#1]}
                \def\@psfile{#1}
}
\def\@setpsheight#1{
                \@heighttrue
                \@darg=#1\relax
                \edef\@psheight{\number\@darg}
}
\def\@setpswidth#1{
                \@widthtrue
                \@darg=#1\relax
                \edef\@pswidth{\number\@darg}
}
\def\@setpsscale#1{
                \@scaletrue
                \def\@pshscale{#1}
                \def\@psvscale{#1}
                \@bblen=#1pt\relax
                \@bblen=1000\@bblen
                \def\@texhscale{\expandafter\remove@dim\the\@bblen}
                \let\@texvscale=\@texhscale
}
\def\@setpshscale#1{
                \@scaletrue
                \def\@pshscale{#1}
                \@bblen=#1pt\relax
                \@bblen=1000\@bblen
                \def\@texhscale{\expandafter\remove@dim\the\@bblen}
}

\def\@setpsvscale#1{
                \@scaletrue
                \def\@psvscale{#1}
                \@bblen=#1pt\relax
                \@bblen=1000\@bblen
                \def\@texvscale{\expandafter\remove@dim\the\@bblen}
}

\def\@setparms#1=#2,{\@nameuse{@setps#1}{#2}}
\def\ps@init@parms{
                \@heightfalse \@widthfalse
                \no@bbfalse
                \def\@psbbllx{}\def\@psbblly{}
                \def\@psbburx{}\def\@psbbury{}
                \def\@psheight{}\def\@pswidth{}
                \def\@pshscale{1}\def\@psvscale{1}
                \def\@texhscale{1000}\def\@texvscale{1000}
                \def\@psfile{}
                \def\@sc{}
}
\def\parse@ps@parms#1{
                \@for\@epsfile:=#1\do
                   {\expandafter\@setparms\@epsfile,}}
\newif\ifnot@eof
\newread\ps@stream
\def\bb@search{
        \openin\ps@stream=\@psfile
        \no@bbtrue
        \not@eoftrue
        \catcode`\%=12\relax
        \ifeof\ps@stream\typeout{epsf: File not found}\fi
        \loop
                \read\ps@stream to \line@in
                \global\toks200=\expandafter{\line@in}\relax
                \ifeof\ps@stream \not@eoffalse \fi
                \@bbtest{\toks200}\relax
                \if@bbmatch\not@eoffalse\expandafter\bb@cull\the\toks200\fi
        \ifnot@eof \repeat
        \catcode`\%=14
}       

\catcode`\%=12
\newif\if@bbmatch
\def\@bbtest#1{\expandafter\@a@\the#1
\long\def\@a@#1
        \ifx\@bbtest#2\@bbmatchfalse\else\@bbmatchtrue\fi}
\def\bb@cull 
        \@ifnextchar\space{\@latexbug}{\bb@extract}}
\def\bb@extract #1 #2 #3 #4 {
        \message{BoundingBox: (#1bp,#2bp)--(#3bp,#4bp)}
        \@darg=#1 bp\edef\@psbbllx{\number\@darg}
        \@darg=#2 bp\edef\@psbblly{\number\@darg}
        \@darg=#3 bp\edef\@psbburx{\number\@darg}
        \@darg=#4 bp\edef\@psbbury{\number\@darg}
        \no@bbfalse
}
\catcode`\%=14

\def\compute@bb{
                \bb@search
                \ifno@bb \typeout{epsf: No BoundingBox}
                \stop
                \else
                \@arga=\@psbburx
                \advance\@arga by -\@psbbllx
                \edef\@bbw{\number\@arga}
                \@arga=\@psbbury
                \advance\@arga by -\@psbblly
                \edef\@bbh{\number\@arga}
                \fi
}
\def\in@hundreds#1#2#3{\@argb=#2 \@argc=#3
                     \@ctmpa=\@argb     
                     \divide\@ctmpa by \@argc
                     \@ctmpb=\@ctmpa
                     \multiply\@ctmpb by \@argc
                     \advance\@argb by -\@ctmpb
                     \multiply\@argb by 10
                     \@ctmpb=\@argb     
                     \divide\@ctmpb by \@argc
                     \@ctmpc=\@ctmpb
                     \multiply\@ctmpc by \@argc
                     \advance\@argb by -\@ctmpc
                     \multiply\@argb by 10
                     \@ctmpc=\@argb     
                     \divide\@ctmpc by \@argc
                     \@arga=#1\@ctmpe=0
                     \@ctmpd=\@arga
                        \multiply\@ctmpd by \@ctmpa
                        \advance\@ctmpe by \@ctmpd
                     \@ctmpd=\@arga
                        \divide\@ctmpd by 10
                        \multiply\@ctmpd by \@ctmpb
                        \advance\@ctmpe by \@ctmpd
                     \@ctmpd=\@arga
                        \divide\@ctmpd by 100
                        \multiply\@ctmpd by \@ctmpc
                        \advance\@ctmpe by \@ctmpd
                     \edef\@result{\number\@ctmpe}
}
\def\compute@wfromh{
                \in@hundreds{\@psheight}{\@bbw}{\@bbh}
                \edef\@pswidth{\@result}
}
\def\compute@hfromw{
                \in@hundreds{\@pswidth}{\@bbh}{\@bbw}
                \edef\@psheight{\@result}
}
\def\compute@handw{
        \if@height 
                \if@width
                \else
                        \compute@wfromh
                \fi
        \else 
                \if@width
                        \compute@hfromw
                \else
                        \if@scale
                                \in@hundreds{\@texvscale}{\@bbh}{1000}
                                \let\@bbh=\@result
                                \in@hundreds{\@texhscale}{\@bbw}{1000}
                                \let\@bbw=\@result
                        \fi
                                \edef\@psheight{\@bbh}
                                \edef\@pswidth{\@bbw}
                \fi
        \fi
}
{\catcode`\p=12\catcode`\t=12
\gdef\remove@dim#1.#2pt{#1}}

\def\compute@sizes{
        \compute@bb
        \compute@handw
}
\def\epsfile#1{
        \ps@init@parms
        \parse@ps@parms{#1}
        \compute@sizes
        \@arga=\@psheight
        \divide\@arga by 65536
        \edef\@psvsize{\number\@arga}
        \@arga=\@pswidth
        \divide\@arga by 65536
        \edef\@pshsize{\number\@arga}
        \message{=>(\@pshsize bp,\@psvsize bp)}
        \leavevmode
        \vbox to \@psheight true sp{
                \hbox to \@pswidth true sp{
                \ifepsfdraft\hss\@psfile\hss\else
                \if@height 
                        \if@width
                                \special{epsfile=\@psfile \space 
                                hsize=\@pshsize \space
                                vsize=\@psvsize \space}
                        \else
                                \special{epsfile=\@psfile \space 
                                vsize=\@psvsize \space}
                        \fi
                \else 
                        \if@width
                                \special{epsfile=\@psfile \space 
                                hsize=\@pshsize \space}
                        \else
                                \if@scale
                                        \special{epsfile=\@psfile \space
                                        vscale=\@psvscale \space
                                        hscale=\@pshscale \space}
                                \else
                                        \special{epsfile=\@psfile \space}
                                \fi
                        \fi
                \fi
                \hfil\fi
                }
        \vfil
        }
}

\catcode`\@=12

\begin{document}

\noindent
{\bf
MAGNETIC AND TRANSPORT PROPERTIES OF 
(La,Sr)MnO$_3$\footnote{to be published in Physica C}
}

\vspace{5mm}

\par\noindent
Nobuo FURUKAWA

\vspace{5mm}

\par\noindent
 Institute for Solid State Physics,
  University of Tokyo, Roppongi 7-22-1,
  Minato-ku, Tokyo 106, Japan

\vspace{5mm}

\par\noindent
Magnetic and transport properties of the
perovskite-type $3d$ transition-metal oxide \\
(La,Sr)MnO$_3$
are theoretically studied 
using the double-exchange model
in infinite dimension.
Magnetoresistance properties as well as the magnetic transition
temperatures are  in  good agreement with the experimental data.

\vspace{5mm}

\section{INTRODUCTION}

After the discovery of the high-$T_{\rm c}$ superconducting oxides,
many strongly correlated $3d$ electron systems have been
reinvestigated.
Recently, perovskite-type manganese oxides ($R$,$A$)MnO$_3$ 
have been studied extensively.
The above group of materials
exhibit colossal magnetoresistance (MR)
at the carrier doped region, which is interesting
not only from the standpoint of strongly correlated systems
but also from application.
Under appropriate hole doping,
the system becomes a 
ferromagnet which is explained by a
double-exchange mechanism \cite{Zener51,Anderson55}.

Transport properties for filling-controlled
single crystals of {\LSMO} have been 
investigated systematically  \cite{Tokura94,Urushibara95}. 
It has been made clear that the resistivity is
controlled by the magnetization of the system.
In either case of applying external magnetic field or
lowering the temperature below the Curie point,
the resistivity behaves universally as a function of
induced or spontaneous magnetization.
In the small magnetization region,
the universal scaling function is given by
$  {\rho(M)}/{\rho(0)} = 1 - C ({M}/{M_{\rm sat}})^2$,
where $\rho(M)$ is the resistivity and $\rho(0)$ is its zero-field value.
Here $M$ is the magnetization while $M_{\rm sat}$ is its
saturated value.
This experimental fact shows that the magnetism and the
transport properties are strongly correlated.
The experimental data also show that 
the coefficient is $C \sim 4$ at $x \sim 0.175$, 
and the value of $C$ decreases  as the hole concentration $x$
is increased. 

In this paper, we investigate the mechanism of the
colossal MR responces in this family of manganese oxides.

\section{MODEL AND RESULTS}

In this paper, we study the
 double-exchange model 
(or the Kondo lattice model with ferromagnetic spin couplings)
as a microscopic
model for manganese oxides.
We assume that the $3d$ electrons in $t_{2\rm g}$ orbitals form
localized spins while electrons in $e_{\rm g}$ orbitals form
itinerant band. Localized spins and band electrons are
strongly coupled with Hund's interaction.
The Hamiltonian is described as
\bequ
  \Ham = 
  - t \sum_{<ij>,\sigma}
        \left(  c_{i\sigma}\dags c_{j\sigma} + h.c. \right)
    -J \sum_i \vec \sigma_i \cdot \vec m_i,
    \label{HamDXM}
\eequ
where $ \vec m_i = (m_i{}^x, m_i{}^y, m_i{}^z)$ is the
classical spin with normalization $|\vec m|^2 = 1$.
In  the infinite-dimensional limit $D\to\infty$,
 Green's function as well as magnetic transition temperatures
and conductivity are obtained exactly \cite{Furukawa94}.
We consider the Bethe lattice
so that the density of states (DOS) is semicircular with
the bandwidth $W$.
The action of the system has the form
\beqarr
  S(\tilde G_0,\vecn) &=&
  - \int_0^\beta \rmd \tau_1
   \int_0^\beta \rmd \tau_2\ 
     \Psi^* (\tau_1) \tilde G_0^{-1}(\tau_1 - \tau_2) \Psi(\tau_2)
        \nonumber \\
  & &  -J  \int_0^\beta \rmd \tau \ 
      \vecn \cdot \Psi^*(\tau) 
           \vec \sigma  \Psi(\tau),
	\label{Action}
\eeqarr
where $\Psi$ and $\Psi^*$ are the Grassmann variables
and $|\vecn| = 1$. 
Green's function $\tilde G_0$ is the Weiss field which 
contains the information about the electron transfer
and the interaction.

\begin{figure}
\epsfile{file=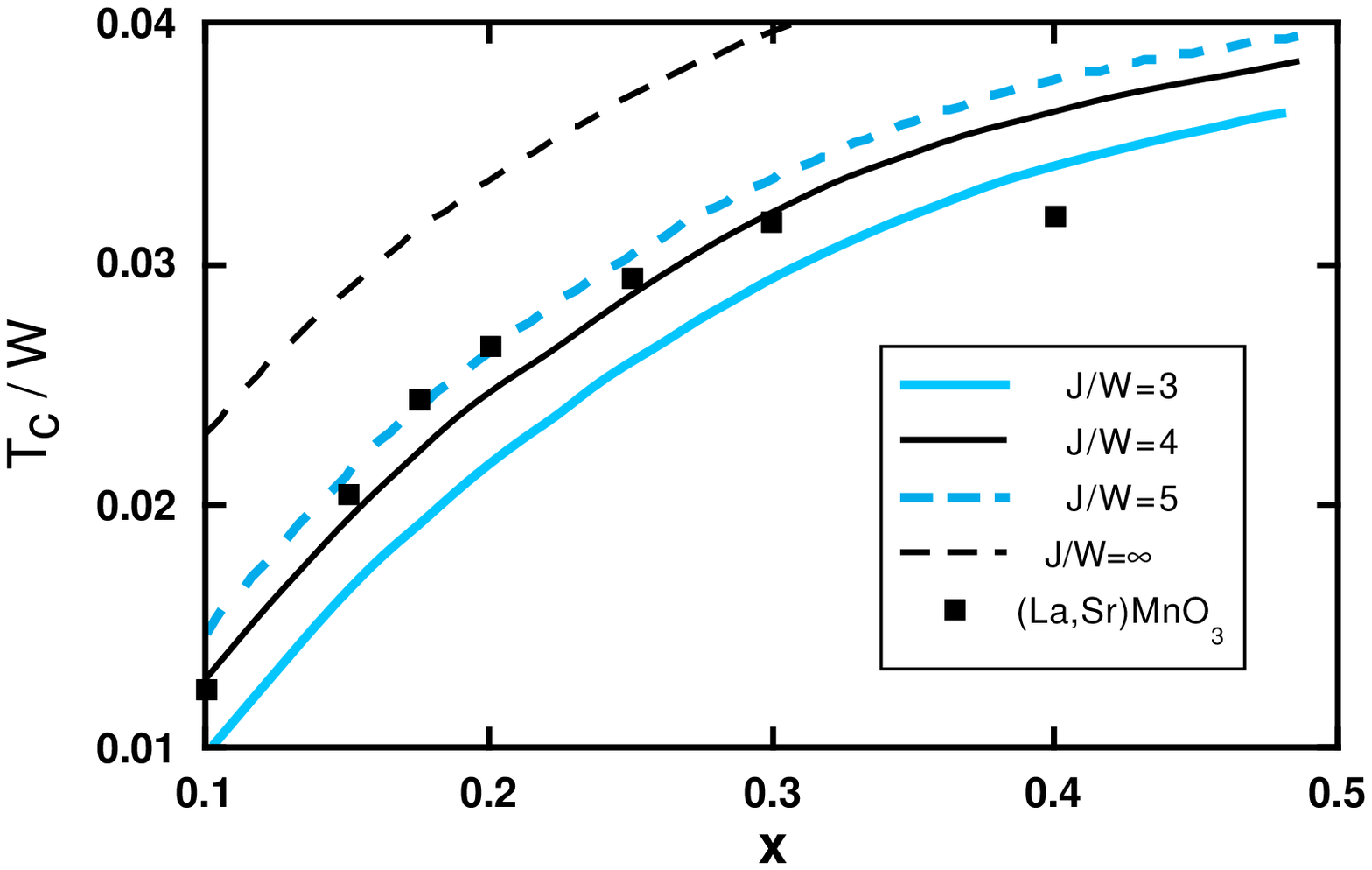,scale=0.96}
\caption{Curie Temperature as a function of doping.}

\vbox to 0pt{
\vspace{-5.1cm}
\hspace{13.3cm}$\infty$
}
\label{FigTc}

\end{figure}

Figure \ref{FigTc} shows the Curie temperature $T_{\rm c}$ as
a function of doping concentration $x$, together with the
experimental data of {\LSMO} \cite{Urushibara95}.
Here we assume $W = 1{\rm eV}$ as a unit of energy,
and the experimental data are also scaled by $W$.
Increase of $T_{\rm c}$ is observed as the hole concentration $x$
is increased due to the increase of the kinetic energy.
For the set of parameters $W\simeq 1 {\rm eV}$ and $J/W \simeq 4$,
we see that the Curie temperatures of {\LSMO} at $0.1 \simle x \simle 0.2$
are well reproduced by the above simplified model \cite{Furukawa95b}.

Now we calculate the MR curve. 
Conductivity  is  
calculated from the Kubo formula.
Since vertex corrections vanish in infinite 
dimension \cite{Khurana90}, we have
\beqarr
  \sigma_{\rm dc}  &=& \sigma_0 W^2
  \sum_\sigma \int N_0(\epsilon) \rmd \epsilon 
\int \rmd \omega 
   \left(-\frac{ \partial f }{\partial \omega} \right)
      A_\sigma{}^2(\epsilon,\omega) ,
   \label{defDCcond}
\eeqarr
where constant $\sigma_0$ gives the unit of conductivity
and $f$ is the Fermi distribution function.
The spectral weight $A_\sigma(\epsilon,\omega)$ is defined by
\beqarr
  A_\sigma(\epsilon,\omega) &=& -\frac1\pi\Im
  \frac1{\omega-(\epsilon-\mu) - \Sigma_\sigma(\omega+\rmi\eta)}.
     \label{defSPfun}
\eeqarr
The resistivity $\rho = 1/\sigma_{\rm dc}$
 and the magnetization $M$ is calculated as a function
of temperature and the external magnetic field.

\begin{figure}
\epsfile{file=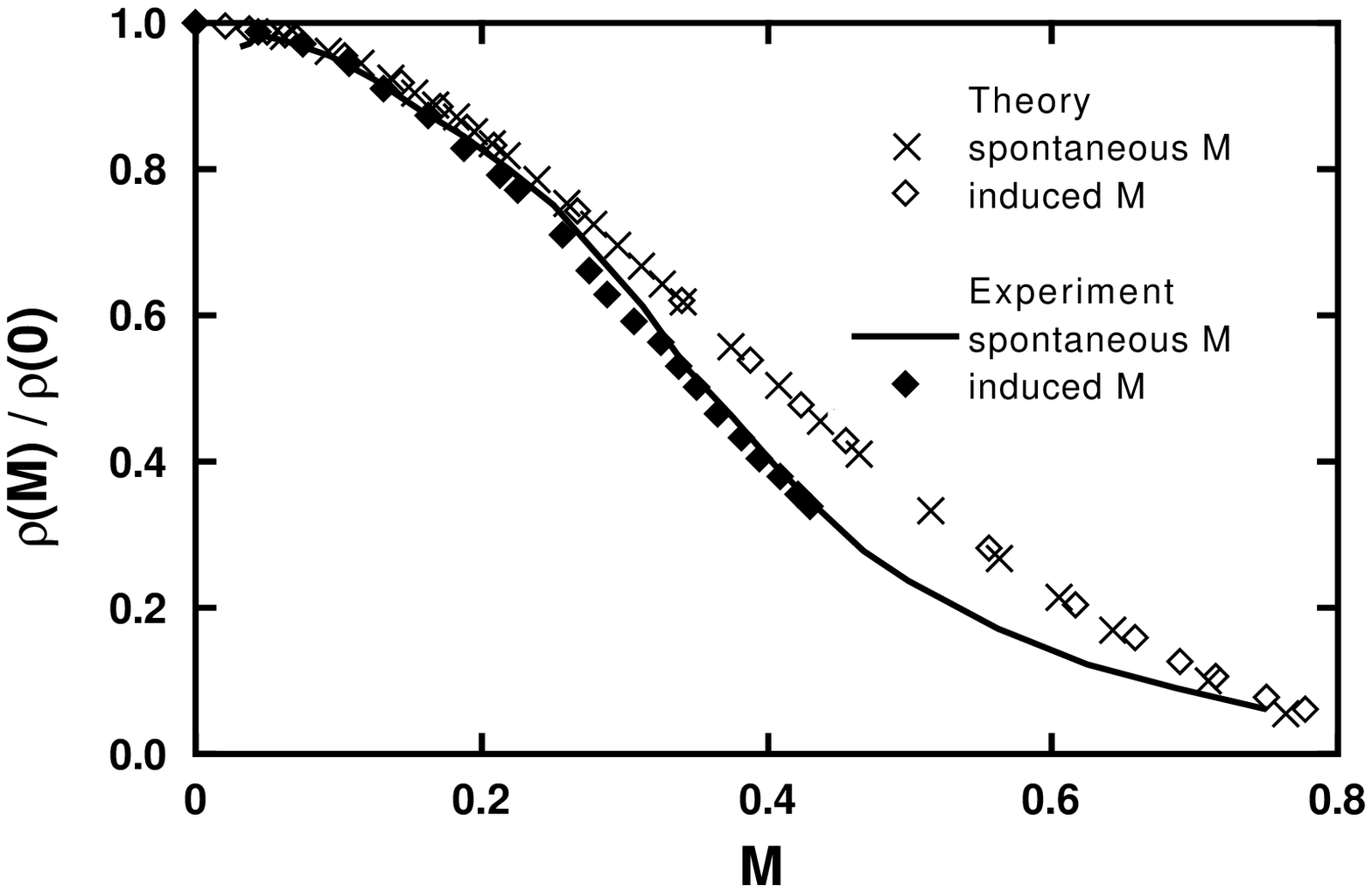,scale=0.96}
\caption{Resistivity as a function of 
 spontaneous magnetization and 
field-induced magnetization.}

\label{FigMR}
\end{figure}

In Fig.~\ref{FigMR},  we show $\rho(M)/\rho(0)$ at $x=0.175$
as a function of magnetization for the cases of applying the external
magnetic field $H$ and changing temperature $T$ below the
Curie temperature. Here we take $J/W=4$.
We  then make a comparison
 with the experimental data of {\LSMO} at $x=0.175$ \cite{Urushibara95}.
In Fig.~\ref{FigMR}, we also plot the resistivity at $T=294{\rm K}$
as a function of induced magnetization normalized by its zero-field value.
Temperature dependence of the resistivity below $T_\rmc$ with
an appropriate normalization is also
shown as a function of spontaneous magnetization.
The experimental result for {\LSMO} is well reproduced
in a quantitative way.
We see that the magnetization is the essential 
thermodynamical variable that determines the MR responce.

The MR curve is obtained analytically if we consider the case
of Lorentzian DOS with $J\to\infty$ \cite{Furukawa95c}.
Using the Kubo formula, we obtain
\bequ
  \frac{ \rho(0) - \rho(M)}{\rho(0)} = 
	\frac{ (4 + 3B) M^2 + BM^4}{1+(3+3B)M^2+BM^4},
\eequ
where
\bequ
   B = \frac{\cos 2\pi x} { 2 - \cos 2\pi x}.
\eequ
The result at $x=0.175$ is in good agreement with the
experimental data of {\LSMO} at $x=0.175$.
At $M \ll 1$, we have
\bequ
 \frac{\rho(M)}{\rho(0)}
 = 1- \frac{ 8 - \cos 2\pi x}{ 2 - \cos 2 \pi x} M^2.
\eequ
We see that $C$ monotonically decreases as $x$ is increased.

\section{DISCUSSION}

Thus we see that the double-exchange model reproduces
the universal curve of MR observed in {\LSMO}.
The {\em mechanism} of the MR as well as the metallic ferromagnetism
are well understood from this minimal model.
We still have to emphasize here that the double-exchange model is
merely an effective model where many kind of degrees of freedom
are reduced. 
In manganese oxides, we have to deal with electrons
in Mn $3d$ $t_{2\rm g}$, $e_{\rm g}$ and O $2p$ orbitals.
We obtain the double-exchange model as a consequence of
neglecting high-energy excitations as well as 
renormalizing Coulomb interactions and lattice effects.
Therefore, we should always regard the model as 
the renormalized model with parameters
$ t= t_{\rm eff}$ and $J = J_{\rm eff}$ in eq.~(\ref{HamDXM}).
Or, in a strict sense, we should consider the action
of the double-exchange model in eq.~(\ref{Action}) with
renormalized Green's function $ G_{\rm eff}$
and the coupling strength $J_{\rm eff}$.

In the strong coupling limit $J\to\infty$,
 the only relevant parameter for
the MR curve in the normalized form $\rho(M)/\rho(0)$  is the carrier 
number $x$.
Therefore, as long as the MR curve is concerned, the double-exchange
model is the relevant model,
since it does not depend on other microscopic 
parameters. 
Of course, such renormalization is only valid
in a certain range of parameters which flow to the
double-exchange fixed point.
Study of such renormalization flow is important in order to specify
what region of A site atom combination in the family of manganese oxides
does the MR response
reproduced by the above simplified double-exchange model.

There are still many thermodynamical quantities that
can not be explained by the double-exchange model alone.
One example is  the absolute value of the resistivity
which should depend on microscopic
 parameters. It is also affected by the 
renormalization of the quasi-particle weight $z$.
In such cases, thermodynamical quantities strongly depend
on interactions such as lattice effects
that are irrelevant to the universality of the model.
Recently, the double-exchange model which also takes into account
the effect of dynamic Jahn-Teller distortion is
studied using the infinite-dimensional approach \cite{Millis95x}.
The result explains the temperature dependence of {\LSMO}
in the region of both above and below the Curie point.
Above the Curie point, the increase of the resistivity
is explained by the effect of Jahn-Teller distortion
while below $T_{\rm c}$ the decrease of the resistivity is mainly
due to the double-exchange mechanism.

The above result may be interpreted as the consequence of
the renormalization of Green's function $\tilde G_{\rm eff}$
by the thermodynamical fluctuation of Jahn-Teller field
together with the double-exchange mechanism.
Therefore, from the point of view of {\em controlling} the MR
response, it is very important to study the renormalization effects from
Coulomb repulsions and lattice distortions.
Especially, the lattice effects  appear in the energy scale
of room temperature.
Further study on the lattice effects as well as Coulomb interactions
are necessary to understand the complex phenomena observed
in manganese oxides.

\section*{Acknowledgments}

This work was supported by a Grant-in-Aid for 
Encouragement of Young Scientists
from the Ministry of Education, Science and Culture.

\bibliographystyle{jpsj}
\bibliography{mymacro,infd,lamno3,local}


\end{document}